\documentstyle[12pt]{article}
\begin{document}
\parindent 2em
\parskip 0ex
\baselineskip 3.5ex
\jot 0.5cm
\tolerance=1500
\flushbottom

\title{Comparison of the Spherical Averaged Pseudopotential Model with the Stabilized 
Jellium Model}
\author{Armando Vieira$^1$, M. Bego\~na Torres$^2$, Carlos Fiolhais$^1$ and L. Carlos 
Balb\'as$^2$\\
\em $^1$Departamento de Fisica da Universidade de Coimbra, \\
\em P-3000 Coimbra, Portugal\\
\em ${^2}$Departamento de Física Teórica, Universidad de Valladolid, \\
\em E-47011 Valladolid, Spain}
\setcounter{page}{1}   
\maketitle   

\vspace{3cm}
 

\clearpage
\begin{abstract}   
We compare Kohn-Sham results (density, cohesive energy, size  and effect of charging) of the 
Spherical Averaged Pseudopotential Model with the Stabilized Jellium Model for clusters of 
sodium and aluminum with less than 20 atoms. We find that the Stabilized Jellium Model, 
although conceptually and practically more simple, gives better results for the cohesive energy 
and the elastic stiffness. We use the Local Density Approximation as well as the Generalized 
Gradient Approximation to the exchange and correlation energies.
\end{abstract}   
\clearpage

\section{Introduction}
The energy functional for an electronic system under the influence of a positive charge 
distribution (ions) is 
\begin{eqnarray} 
E[n,n_+]& =&T[n] +E_{xc}[n]
+{1\over 2} \int d^3 r \int d^3 r'\,{n(\vec{r})
n({\vec{r'}})\over |\vec{r} - {\vec{r'}}|}\\ \nonumber
&+&\int d^3 r\,V_{+}(\vec{r}) n(\vec{r})
+ \frac{1}{2}\int d^3r\int d^3r \, \frac{n_+(\vec{r}) \,n_+(\vec{r'})}{|\vec{r}-
\vec{r'}|}\enskip,
\label{efunt}  
\end{eqnarray} 
where $n(\vec{r})$ is the density of the valence electrons, $n_+(\vec{r})$ is the ionic 
density, $T$ is the kinetic energy, $E_{xc}$ is the exchange and correlation energy, and 
$V_+(\vec{r})$ is the potential created by the ions. Atomic units are used in this paper.

The Jellium Model is a simple model to describe simple metals. To study a spherical 
cluster in this model, the ions are replaced by a continuous positive background which is 
constant inside a sphere of radius $R_{jel}=r_s \, N^{1/3}$, where $N$ is the number of 
valence electrons, and zero outside: 
\begin{equation}
n_+(r)= \overline{n}\,\, \theta(R_{jel}-r) \enskip,
\end{equation}
with $\overline{n}=3/(4\pi r_s^3)$. For this model the external potential is simply
\begin{equation}   
V_+(\vec{r}) \equiv v_{jel}(r) = \left\{\begin{array}{ll} -{N\over 2R_{jel}} \bigg[3 -   
\Big({r\over R_{jel}}\Big)^2\bigg]   
\enskip \enskip & (r < R_{jel}) \\ [.3cm]   
-{N\over r} & (r \geq R_{jel}). \enskip \end{array} \right.   
\label{poten}
\end{equation}   

The Stabilized Jellium Model (SJM) \cite{perdew3} represents  an improvement over the 
Jellium Model. For the bulk matter, it yields more realistic binding energies and bulk 
moduli~\cite{perdew3}. For semi-infinite systems, it provides more correct surface energies and 
work functions~\cite{fiolhais}.  For clusters, it gives very reasonable cohesive energies, 
dissociation energies, ionization energies, etc.~\cite{marta, ziesche1}. Its energy functional  may 
be written as 

\begin{equation} 
\nonumber E_{SJ}[n,n_+]= E_J[n,n_+]+\left<\delta v\right>\int d^3\,r\, \theta(r-
R_{jel})\,[n(r)-n_+(r)]+\tilde{\epsilon} \int d^3 r\,n_+(r)\enskip,
\label{efun}  \end{equation} 
where $E_J[n,n_+]$ is the jellium energy functional (Eq. (1) with the external 
potential given by Eq. (\ref{poten})), and
\begin{equation}
\left<\delta v\right> = \frac{1}{\nu}\int_0^{r_0} dr \,r^2\left[w(r) - v_{jel}(r)\right] =
-\frac{3}{10}\frac{\nu^{\frac{2}{3}}}{r_s}+ \frac{3}{2} \frac{r_c^2}{r_s^3}\enskip
\label{erc}
\end{equation}
is the average, in the Wigner-Seitz cell (taken to be a sphere of radius 
$r_0=r_s\nu^{1/3}$, with $\nu$ the valence), of the difference between the 
pseudopotential 
\begin{equation} 
w(r) = \left\{
\begin{array}{ll}
0    & (r < r_c)\\
-\nu/r & (r \geq r_c) 
\end{array} \right.
\end{equation} 
(Ashcroft empty core potential~\cite{ash}, with core radius $r_c$) and the 
jellium potential. 

In the last term of Eq. (\ref{efunt}), $\tilde{\epsilon}$ is the spurious background repulsion 
inside each Wigner-Seitz cell, which is equal to the sum of the  Madelung energy 
\begin{equation} 
\epsilon_M=\frac{1}{\nu}\int_0^{r_0}dr\,4\pi r^2 \overline{n}\left(\frac{-
\nu}{r}+\frac{1}{2}v_{jel}\right)= -\frac{9}{10}\frac{\nu^{\frac {2} 
{3}}}{r_s}
\end{equation} 
and the average of the pseudopotential repulsion~\cite{perdew3}: 
\begin{equation} 
\frac{1}{\nu}\int_0^{r_c}dr\,4 \pi r^2 
\overline{n}\frac{\nu}{r}=\frac{3}{2}\frac{r_c^2}{r_s^3}\enskip.
\end{equation} 

Physically, Eq. (\ref{efun}) means that we start with jellium and  localize uniformly the 
ions  but their interaction with the valence electrons (accounted for by the pseudopotential)   is 
only taken into account perturbatively and in a spherical averaged way.

With the energy functional (\ref{efun}) we may extract results for clusters, via the 
Kohn-Sham method in the Local Density Approximation (LDA) for exchange and correlation, 
going from the single stabilized jellium atom all the way up to the bulk solid. 
We use the Perdew-Wang interpolative formula~\cite{pcorr} for the correlation energy.
We consider that the jellium background is either fixed with the bulk density value or 
relaxed in order to give the minimal energy for a given number of particles (using a 
fixed pseudopotential, {\it e.g.}, transferred from the bulk). In the latter case, which is in 
principle more realistic, we speak of self-compression of the neutral cluster, a phenomenon 
which may be classically explained by the effect of the surface tension on the 
density~\cite{perdew8, vieira5,vieira2}. We take here, when not otherwise indicated, the  
SJM including self-compression. The energetics of the cluster ground states is similar to 
that of the Jellium Model, with the same shell structure, but shifted down, {\it i.e.}, the 
valence electrons are more bound.

Let $E=E(N,r_s,r_c,\nu)$ be the binding energy of a neutral spherical cluster with $N$ 
valence electrons. The equilibrium ionic density, $r_s^*$, of a cluster is obtained imposing the 
following equilibrium condition:

\begin{equation}
\frac{\partial}{\partial r_s} \left[\frac{E(N,r_s,r_c,\nu)} {N}\right]_{r_s=r_s^*}=0,
\label{61}
\end{equation}
where the derivative is taken at constant $N$ and $r_c=r_c(r_s^B,\nu)$, where $r_s^B$ is 
the bulk density parameter. The core radius  is fixed by the bulk  stability condition, which is Eq. 
(9) in the limit $ N \rightarrow \infty $. We have self-compression if $r_s^*<r_s^B$ 
and self-expansion if $r_s^*>r_s^B$.

The elastic stiffness of the cluster is defined by
\begin{equation}
B(N,r_s^*,r_c,\nu)=-V\;\frac{\partial^2 E}{\partial V^2}|_N=
\frac{1}{12 \pi r_s^* N}
\frac{\partial^2}{\partial r_s^2} 
\left[\frac{E(N,r_s^*,r_c,\nu)}{N}\right]_{r_s=r_s^*}\enskip.
\label{eb}
\end{equation}
In the limit $N\rightarrow \infty$, this quantity goes over to the bulk modulus. Stability 
demands that $B\geq0$.

On the other hand, the Spherical Averaged Pseudopotential Model (SAPS) takes into account 
the geometrical arrangement of the ions located at 
positions $\vec{R}_j$~\cite{saps,saps1}. This is optimized by minimizing the total 
energy of the cluster. 

Only valence electrons are explicitly treated, as in the SJM, being the ion cores replaced 
by atomic pseudopotentials $w(|\vec{r}-\vec{R}_j|)$. The simplification introduced by 
the SAPS model for clusters of simple metals consists of replacing the external potential 
felt  by the valence electrons, $V_+(\vec{r}) = \sum_j^{N_{at}} w(|\vec{r}-\vec{R}_j|)$, by 
its spherical average, $\left<V_+\right>(r)$, around the cluster center. In this way, the energy 
functional in the SAPS model reads as

\begin{eqnarray} 
E_{SAPS}[n,\vec{R}_i]&=&T[n]+E_{xc}[n]+
{1\over 2} \int d^3 r \int d^3 r'\,{n(\vec{r})\,   
n({\vec{r'}})\over |\vec{r} - {\vec{r'}}|}\\ \nonumber
&+&4 \pi r^2 \int dr \left<V_+\right> n(r)+ 
\frac{1}{2}\sum_{i\neq j}^{N_{at}}{\frac{z_i z_j}{|\vec{R}_i-\vec{R}_j|}}\enskip,
\label{esaps}
\end{eqnarray} 
where $z_i$ denotes the charge of the ions and $N_{at}=N/\nu$ is the number of atoms. 
The last term in Eq. (11) represents the repulsion between the ion cores, which is 
approximated by the interaction between point charges. 
As in the SJM, we have used the LDA with the correlation energy functional of 
Perdew-Wang. 

In order to determine the cluster energy, we start from a randomly initial atomic configuration 
$\vec{R}^0_j$, and solve the Kohn-Sham equations, evaluating the valence electron density 
and the total energy given by Eq. (11). To obtain the geometry that minimizes the total energy, 
we calculate the energy variation $(\delta E_{j})_{\alpha}$, corresponding to the displacement 
of the ion at position $\vec{R}_j$ in each Cartesian direction ($\alpha=x, y, z$), by a small 
quantity $\delta d$, keeping the electron density and the other ions coordinates frozen.
Then we choose the maximal energy variation $\delta E_{max}$, and move each ion coordinate 
by the amount $[(\delta E_j)_{\alpha}/ \delta E_{max}]\, \delta d$, obtaining a new geometry, 
$\vec{R}^1_j$, for which we calculate the corresponding electron density.
With this new geometry  we repeat the cycle again, being the process iterated until convergence 
is achieved and an equilibrium geometry is found. This generally leads to a local minimum. To 
find the global minimum we have repeated the calculations for a large number (typically 20) of 
random initial configurations.

To obtain the elastic stiffness in the SAPS model, we use an analogue of Eq. (\ref{eb}) 
with the approximation
\begin{equation}
\frac{\partial^2 E}{\partial r^2}\simeq \frac{E(\vec{R}_j-\delta \vec{r}) + 
E(\vec{R}_j+\delta \vec{r})- 2 E(\vec{R}_j)}{ |\delta r|^2}\enskip.
\end{equation}
The energy of the cluster is calculated with the atoms at the equilibrium positions 
$\vec{R}_j$ and displaced radially in and out by a small amount, $\delta \vec{r}$, from 
$\vec{R}_j$ (for simplicity we have omitted the energy functional dependence on $n$).

An extension of the SAPS model which exploit all ionic degrees of freedom in three 
dimensions but restricts the electronic many-body problem to axial symmetry has been 
recently introduced by Montag and Reinhard \cite{montag94,montag95}. This model is 
known as CAPS (Cylindrical-Averaged Pseudopotential Scheme), since a cylindrical average is 
taken instead of a spherical average.

In this work we compare the SAPS model with the SJM because the two schemes use a 
spherical average of the atomic pseudopotential, and one could expect the two descriptions of 
some clusters properties to be similar. In particular, one could expect  that the 
phenomena of self-compression of  neutral clusters with respect to the bulk density and 
self-compression or self-expansion of charged clusters, which have been identified in the 
SJM~\cite{vieira5,vieira2}, are also described  in the SAPS. With this goal, we examine the 
two models for small clusters of sodium and aluminum. In the Appendix we compare the LDA 
cohesive energies with the Generalized Gradient Approximation (GGA) ones.

\section{Results}
Fig. 1 shows the SAPS valence electron density for 8-atoms clusters of sodium 
($r_s=3.93$, $\nu=1$) and aluminum ($r_s=2.07$, $\nu=3$), in comparison with the SJM 
one, if the jellium edge is allowed to move. In the SAPS, we have used the recently 
proposed evanescent core pseudopotential model, which has the advantages of being local and 
analytical and which describes generally well the main physical properties of bulk simple 
metals~\cite{fiolhais1,fn1,fn2}. In the SJM, we have used the simpler and more common 
Ashcroft pseudopotential with the core radius fixed by the condition of bulk stability. We 
could as well have used any other local pseudopotential with a single free parameter, 
arriving at the same results. We conclude that the two densities are somewhat different, 
especially in the interior for aluminum (the SAPS density is much lower there). 
The surface diffusivity is bigger for SAPS than for SJM (this leads to a larger redshift of the 
surface plasmon resonance with respect to the Mie peak~\cite{brack}).
However, the radial probability density $r^2 n(r)$, represented in the inset, is similar in the two 
models. We have performed the same comparison for positively  ionized  systems 
(Na$_8^{3+}$ and Al$_8^{12+}$), arriving at the same conclusions.

Let us  consider the energy of the cluster in the two models. The cohesive energy is the 
difference between the energy of the free atom and the energy of the cluster per atom, 
\begin{equation} 
E_{coh}(N_{at})= E(\nu)-\frac{E(\nu N_{at})}{N_{at}} \enskip.
\label{13}  \end{equation} 
For the bulk, Eq. (\ref{13}) leads to the heat of sublimation: $E_{coh}(N_{at}=\infty)=E(\nu)-
a_v \nu$, where $a_v$ is the bulk 
energy per valence electron. Fig. 2 represents the cohesive
energy for small clusters of sodium and aluminum 
using the SAPS model, and the two versions of the SJM: one with the self-compression 
effect and another with the background density frozen at the bulk value,
 $r_s^B$. In Fig. 2 we have also plotted the liquid drop energy in the SJM
\begin{equation}
E_{coh}^{LDM}(N_{at}) = a_s \nu^{2/3}\,(1-N_{at}^{-1/3}) + a_c \nu^{1/3}\,(1-
N_{at}^{-2/3})\enskip,
\label{}  \end{equation}
with $a_s$ and $a_c$ the surface and curvature energy coefficients. In 
the case of self-compression, the curvature energy coefficient is replaced 
by~\cite{perdew8}
\begin{equation}
\tilde{a}_c=a_c-\frac{1}{2}\frac{(a_s')^2}{a_v''}\enskip.
\end{equation}
For sodium we have $\tilde{a}_c=0.18$ 
eV, and for aluminum $\tilde{a}_c=-0.10$ eV, which contrast with the 
non-compression values, respectively 0.26 eV and 0.65 eV. The surface energy ($a_s=0.57$ 
eV for sodium and 0.86 eV for aluminum) is not affected by compression.

The cohesive energy in the SAPS is much lower than any of the SJM results, although a 
similar pattern (with maxima at shell closures) is apparent. 
For sodium, the self-compressed SJM results are in very good agreement with LDA
Car-Parrinello calculations~\cite{andreoni} (and also close to the experimental values 
~\cite{brechignac8}). For aluminum, the frozen density SJM results are in good agreement with 
Car-Parrinello calculations~\cite{jones} (and experiment~\cite{ray}), while self-compressed 
SJM and SAPS yield negative values for very small clusters. This fact should be seen as a 
drawback of the spherical approximation used in both models. 
The spherical restriction, valid for the atom, is indeed an artificial constraint for the bulk solid. 
This makes the SAPS valence-electron energies too high and the cohesive energies too low. In 
the SJM, the errors made for the atom and the solid seem to be similar.

The SAPS result for the cohesive energy depends somewhat on the selected 
pseudopotential, but replacing the evanescent core pseudopotential by a different one such 
as Manninen's~\cite{maninen}, the SAPS results remain negative for small clusters. 
The error made in using the pseudopotential for the atom may be estimated replacing the 
binding energy of the pseudo-atom by the sum of the first $\nu$ ionization potentials in an 
all-electron calculation using the same exchange and correlation energies. That correction, 
using the evanescent core pseudopotential, yields a shift upwards in the cohesive energy 
of 0.2 eV  for sodium, and 0.0 eV for aluminum.

In Appendix, we compare the LDA with the GGA results for the cohesive energy, concluding 
that the density gradient corrections reduce this quantity.

Besides the electronic shell structure, leading to magic numbers, which appears in both 
the SJM and the SAPS model, there is in the latter an ionic structure which also reflects itself in 
oscillations of the total energy. In order to disentangle the electronic and geometric shell effects, 
we have evaluated the kinetic energy contribution by means of an Extended Thomas-Fermi 
energy functional with fourth-order gradient corrections (TFDGW4)~\cite{engel,engel1} using, 
however, the self-consistent Kohn-Sham density of each model (Fig. 3). In this way the 
electronic shell-structure due to the quantum kinetic energy operator of the Kohn-Sham 
equations is partially erased.
In both cases, it is clear that with the new functional we would obtain self-consistently a 
different ionic background and ionic structure, but we have just evaluated a semi-classical kinetic 
energy with the quantal density as in Ref. \cite{engel1}. The resulting binding energies of 
sodium neutral clusters are represented in Fig. 3. In the SJM, we obtain a smooth 
curve, which may be fitted by a liquid drop formula; the surface and curvature coefficients 
are $a_s=0.52$ eV and  $a_c=0.22$ eV, which agree very well with results obtained from 
the planar surface problem, respectively 0.58 eV and 0.26 eV~\cite{fiolhais}. But, in the SAPS, 
we are left with some wiggles, which are mainly due to the reorganization of ionic shells. An 
increase of the energy arises when a new shell is added. For instance, this happens when 
going from $N_{at}=4$ to $N_{at}=5$ or from $N_{at}=13$ to $N_{at}=14$. Note that, 
except for the $N_{at}=1$ case, the SJM results is always above the SAPS results.

For the sake of comparing  the cluster size evolution in the two models, we have plotted, 
in Fig. 4, the radius of the outer ionic shell $R_{out}$ in SAPS divided by $R(N)= R_{jel}(N)-
d/2$, where $d$ is the distance between parallel closed packed planes in the bulk solid, against 
the number of atoms. In fact, the planar surface is the limit of the curved surface of a big cluster, 
and it is known from surface physics that the first lattice plane is located at a distance $x=- d/2$ 
from  the jellium edge located at $x=0$.
The quantity $d =1.436\, r_0 $ is the distance between the (110) planes in the bcc 
structure (these are the planes with the biggest separation) and $d= 1.477\, r_0 $ is the 
distance between the (111) planes of the fcc structure. We see  that $R_{out}/ R$ approaches 
1 from above, although it becomes increasingly  difficult  to apply the SAPS to very big clusters 
and to attain the planar surface limit. This dilatation effect of the outer ion shell with respect to 
the surface limit is particularly clear for very small clusters. On the other hand, we observe 
contraction in the SJM: in Fig. 4, $r_s^*/r_s^B$ goes from below to 1 when $N_{at}$ 
increases. In both models, the approach to the asymptotic limit is more rapid for sodium than 
for aluminum.

Notwithstanding the expansion of the outer ionic shell,  the distances between next 
neighbour ions inside each shell are smaller than in the bulk environment differing from 
ionic shell to ionic shell. This inhomogeneous contraction observed in SAPS has been 
reported before~\cite{angel91,glo92}. Such an effect is outside the scope of the SJM, 
since the jellium background must have a constant density~\cite{perdew3}. 

To analyse the behaviour of charged systems we may keep the size fixed ({\it e.g.}, 
Na$_8$ and Al$_8$) and increase the charge all the way up to Coulomb explosion of the 
volume (Fig. 5). In the first case, the radius of the jellium sphere is increasing, while in the SAPS 
the same happens with the radii of the ionic shells.
We see a strong similarity of the phenomenon of volume 
explosion described by the SJM and the SAPS model. The maximal positive charge that 
Na$_8$ can hold is $q=3$, whereas for Al$_8$ is $q=13$ for the SJM and $q=14$ for the 
SAPS (the charged systems of Fig. 1 are therefore close to their stability limit). For Al$_8^+$, 
we observe in both models a small self-compression 
instead of self-expansion with respect to the neutral system. This is an interesting 
effect since it goes against the expectations.

The type of instability  we are describing is different from usual fission since, in the 
latter, volume is fixed while shape is deformed and here volume changes keeping the shape. The 
fissibility, defined as $x = E_C / 2 E_S$, where $E_C$ is the Coulomb 
and $E_S$ is the surface energy of a spherical cluster, controls the fission probability.
For $x=1$ the fission barrier vanishes and the cluster decays spontaneously. 
A different parameter should be considered for the explosion of the background. Anyway, 
within the SJM,
we get $x=3.4$ for Na$_8$ and $x=26.6$ for Al$_8$ as critical fissibilities for volume 
explosion, which indicates that excited clusters prefer to lower their energy by deforming 
and breaking into two or more pieces than by exploding the volume.

Finally, in Fig. 6 we compare the  elastic stiffness of charged aluminum 
clusters. The two values agree when the charge is big, but differ for small charges: the 
elastic stiffness of neutral aluminum clusters in the SAPS model is much larger than in the SJM. 
We remind that the aluminum bulk modulus given by the SJM is a factor of 2 bigger than the 
experimental value, as indicated in the picture.

\section{Conclusions}
We have examined  comparatively some physical properties of small clusters of sodium 
and aluminum as given by the SJM and the SAPS model.

The radial electronic probabilities are similar for sodium and somewhat different for aluminum. 
Although the SJM is conceptually and practically more simple, its results for the cohesive energy 
are in better agreement with more involved theoretical results.  The restriction to the spherical 
symmetry 
seems to be a serious drawback of the SAPS model. The outer ionic radius in the SAPS and the 
jellium edge 
in the SJM for neutral clusters approach the asymptotic limit in different ways. However, the 
results of the two models compare very well when we normalize the ionic radius of a charged 
cluster to that of the corresponding  neutral cluster, and keep the number of atoms fixed when 
increasing the charge. The elastic stiffness given by the SJM, although too high in comparison 
with data for high density metals such as aluminum, is better than the SAPS output.
This indicates that the spherical restriction to the ionic structure, besides giving 
too low cohesive energy for high density metal clusters, also makes them too hard.
Therefore, the corresponding monopole compression mode in the SAPS has an energy which is 
expected to be too high in comparison with experiment~\cite{mananesm}.

In conclusion, the results obtained with the SAPS for the energetics and compression properties 
of small clusters should be taken with some care. Simpler models, such as SJM, may surprisingly 
perform better. On the other hand, the SJM is useless for heterogeneous clusters, while SAPS is 
able to describe them. The two types of models are useful for understanding the main trends of 
small and large clusters, since {\it ab initio} theoretical models are computationally more 
elaborated and, at present, almost impossible to apply to clusters with more than 20 atoms. 
Since the SJM and the SAPS model assume spherical symmetry, they are particularly suitable 
for systems with spherical geometries (for instance, $N_{at}=8$ and 20 for sodium). They are 
not adequate for systems with $N_{at}\leq 5$, which are known, from first principles 
calculations, to be planar.

The assumptions of stabilized jellium are not restricted to a compact spherical shape 
making it more versatile than the SAPS model to describe some properties of
metallic clusters. A simple modification of the SJM which should bring it to a better agreement 
with SAPS is the hollow SJM~\cite{membrado,yanouleas}. Indeed, for a small system, we may 
open a hole in the middle and vary the size of the system so that the energy is minimal for a given 
number of particles. Deformed jellium \cite{brackh} and CAPS~\cite{montag94,montag95} 
represent  improvements of respectively spherical SJM and SAPS which correct the spherical 
approximation.

%**
\section*{Appendix: The Generalized Gradient Approximation for small clusters}
The Local Density Approximation (LDA) for exchange and correlation is the most used scheme 
in Density Functional Theory, but other approximations, which try to correct some of its 
deficiencies ({\it e.g.}, incapacity to bound extra electrons), may be implemented.
One of the most popular is a semi-local approach named Generalized Gradient Approximation 
(GGA) from which there are a couple of versions, {\it e.g. } Refs.~\cite{gga1} 
and~\cite{gga2}. Instead of LDA, we have used the GGA, in the form recently proposed by 
Perdew, Burk and Ernzerhof~\cite{gga2}, for calculating the cohesive energies for sodium and 
aluminum clusters in the SJM as well as in the SAPS. In the SAPS we have taken the ionic 
configuration optimized within  the LDA. For SJM we are not considering self-compression. 

The results are shown in Fig.~\ref{ggana}, in comparison with LDA theory and experimental 
data. We obtain a small decrease of the cohesive energy for both metals. This fact reiterates our 
conclusion, reached with the LDA, that the SJM is in better agreement with the experimental 
data than the SAPS. The decrease of the cohesive energy, due to gradient corrections, is a well-
known feature of Density Functional Theory~\cite{gga1}.

\vspace{.5cm}  
{\bf Acknowledgments}\\  
We are indebted to J.P. Perdew (Tulane University, USA) and A. Ma\~nanes (Universidad 
de Cantabria) for their helpful comments. This work has been partially supported by the 
Praxis XXI projects No. 2/2.1/FIS/26/94 and No. 2/2.1/FIS/473/94. One of us (A. V.) has 
been supported by a grant of the Praxis XXI program.

\clearpage

%\section*{List of captions for figures}

\begin{figure}
\caption{Valence electron densities of neutral and charged octamers of sodium and 
aluminum, 
obtained with the SAPS and the SJM. In insert, we represent the radial probability 
density of the neutral clusters, $r^2 n(r)$, in arbitrary units. The horizontal line 
represents the background
density of the neutral cluster in the SJM (its radius is $R_{jel}= r_s^* N^{\frac {1} 
{3}}$, with $r_s^*= 3.70$, for sodium, and $r_s^*=1.92$ for aluminum.
The vertical arrows on the horizontal axis denote the position of ionic 
shells in the SAPS: in each picture the left arrow marks the position of the ionic shell 
for the 
neutral cluster, whereas the right one stands for the charged case. Each cluster has a 
single ionic shell, 
with D$_{4d}$ symmetry.} 
\label{dens}\end{figure}

\begin{figure} 
\caption{Cohesive energies of sodium and aluminum clusters obtained in the SAPS 
model and the SJM. The SJM is shown in two versions: one, denoted by 
SJM$(r_s^B)$, with the background density fixed at the bulk value, and another 
denoted by SJM$(r_s^*)$ including 
self-compression, where the background density of the cluster is allowed to change. 
The solid lines are obtained with the liquid drop formula, with (below) and without 
(above) the 
self-compression effect. (We take the Kohn-Sham energy of the atom instead of the liquid drop 
value.) Ou results are compared with Car-Parrinello calculations~\protect\cite{andreoni,jones}} 
\label{ecoes}
\end{figure} 

\begin{figure} 
\caption{Binding energies per atom for sodium clusters in the SJM and SAPS models. 
The lines marked with TFDGW4 represents the energy obtained by replacing the quantal kinetic 
energy by the TFDGW4 functional of the Kohn-Sham density. The numbers denote the 
structure of ionic shells in SAPS, and have the following meaning: 1- one ionic shell; 2- one ionic 
shell with one atom at the center; 3- two ionic shells; and 4- two ionic shells with one atom at 
the center.}
\label{etf}\end{figure}

\begin{figure} 
\caption{Size evolution of the outer ionic shell radius, $R_{out}$, in the SAPS,  and 
of the equilibrium ionic density of SJM, $r_s^*$, for sodium clusters and 
aluminum. Both quantities are normalized to the corresponding limit $ N \rightarrow
\infty $. The numbers over the SAPS points have the same meaning as in Fig. 3.} 
\label{compN}\end{figure}

\begin{figure} 
\caption{Size of Na$_{8}^{q+}$ and Al$_8^{q+}$, normalized to the corresponding 
size of the neutral system, as a function of charge $q$. For the SAPS we represent the 
ratio between the outer ionic shell radius of the charged and the neutral cluster. For 
the SJM we represent the ratio between the equilibrium density parameter, $r_s^*$, 
of the charged and the neutral cluster. The inset is a blow-up between $q=-2$ and 
$q=3$ (for Al$_8^q$ the  SAPS has no bound solutions for $q<0$).} 
\label{comprs}\end{figure} 

\begin{figure} 
\caption{Elastic stiffness $B$ of Al$_{8}^{q+}$, as a function of charge $q$, 
obtained with the SAPS and the SJM.}
\label{compb}\end{figure}

\begin{figure} 
\caption{Cohesive energies of sodium ({\bf above}) and aluminum ({\bf below}) clusters 
obtained within the SAPS and the SJM, with the LDA and GGA for exchange and correlation.}
\label{ggana}\end{figure}

\end{document}